\documentclass[preprint,authoryear,12pt]{elsarticle}

\usepackage{amssymb}
\usepackage{graphicx}
\usepackage{rotating}        %
\usepackage{algorithmic}  %
\usepackage{algorithm}      %
\usepackage{rotating}

\journal{Social Networks}

\usepackage{color}

\begin{document}

\begin{frontmatter}

\title{Logistic Network Regression for Scalable Analysis of Networks with Joint Edge/Vertex Dynamics\tnoteref{thanks}}

\author[rvt]{Zack W.~Almquist\corref{cor1}} \ead{almquist@uci.edu}
\author[rvt,focal]{Carter T.~Butts} \ead{buttsc@uci.edu}
\cortext[cor1]{Corresponding author} 
\address[rvt]{Department of Sociology, University of California, Irvine,  3151 Social Science Plaza A, Irvine, CA 92697-5100} 
\address[focal]{Institute for Mathematical Behavioral Sciences, University of California, Irvine}

\tnotetext[thanks]{This work was supported in part by ONR award N00014-08-1-1015 and National Science Foundation (NSF) award BCS-0827027.}

\begin{abstract}
Network dynamics may be viewed as a process of change in the edge structure of a network, in the vertex set on which edges are defined, or in both simultaneously. Though early studies of such processes were primarily descriptive, recent work on this topic has increasingly turned to formal statistical models. While showing great promise, many of these modern dynamic models are computationally intensive and scale very poorly in the size of the network under study and/or the number of time points considered.  Likewise, currently employed models focus on edge dynamics, with little support for endogenously changing vertex sets.  Here, we show how an existing approach based on logistic network regression can be extended to serve as highly scalable framework for modeling large networks with dynamic vertex sets. We place this approach within a general dynamic exponential family (ERGM) context, clarifying the assumptions underlying the framework (and providing a clear path for extensions), and show how model assessment methods for cross-sectional networks can be extended to the dynamic case.  Finally, we illustrate this approach on a classic data set involving interactions among windsurfers on a California beach.

\end{abstract}


\end{frontmatter}


\section{Introduction}
\label{intro}

Change in network structure (i.e., network dynamics) has long been a topic of both theoretical and methodological interest within the social network community.  Network dynamics may be viewed as a process of change in the edge structure of a network, in the vertex set on which edges are defined, or in both simultaneously.  While early studies of such processes were primarily descriptive \citep[e.g.,][]{sampson68,newcomb53,coleman64}, recent work on this topic has increasingly turned to formal statistical models \citep[e.g.,][]{banks96, snijders96, snijders01,snijders05,robins01,krackhardt07}.  While showing great promise, many of these modern dynamic models are computationally intensive and scale very poorly in the size of the network under study, making them difficult or impossible to apply to large networks in practical settings.  Likewise, currently employed models focus on edge dynamics, with little support for endogenously changing vertex sets.  Given this situation, there is a need for scalable approaches that -- even if limited in various ways -- can serve as a starting point for analysis of intertemporal network data at large scales.  This paper explores the use of the well-known logistic network regression framework as a simple basis for the modeling of joint edge/vertex dynamics with various orders of temporal dependence.  We expand on past work showing how this family can be derived from the theory of Exponential Family Random Graph Models (ERGMs) \citep{holland81a, holland81b, butts08, snijders02,strauss} via dependence assumptions in the dynamic case, and discuss computational issues related to its use with large, sparse graphs.  We discuss basic parameterization issues, including one approach to the treatment of cases with vertex set dynamics. We follow this discussion with a case study in which we analyze the dynamics of interpersonal communication during 31 days of windsurfer interaction on a beach in Southern California, the famous ``beach" data-set collected by \cite{freeman88} (hereon referred to as the \emph{beach network}). Demonstrating several methods for assessing model adequacy, we evaluate the ability of the logistic family to capture the evolution of the beach network over the 31 day collection period. Informed by these results, we conclude by discussing some of the strengths and weaknesses of this approach for practical analysis of large-scale intertemporal data sets. 

Although existing models for joint edge/vertex evolution are rare \citep[an example being recent work by][]{krivitsky}, basic statistical methods for edge prediction have been in the social network literature for several decades \citep[see, e.g.][]{krackhardt87a,krackhardt87b,krackhardt88}. Much of this early work involved variations on OLS or logistic regression applied to adjacency matrices.  Logistic regression \emph{per se} has a long history of being applied to social network data \citep{robbins99,wasserman96,pattison99,lazega97}, due both to the fact that it arises naturally from edgewise independence assumptions \citep[see][]{holland81a,holland81b} and to the wide availability of existing implementations.  Less appreciated have been the computational advantages of the logistic framework relative to more complex schemes; methods for estimation of logistic models on large, sparse data sets are well-developed \citep[see, e.g.][]{komarek03,komarek04,lin08}, in contrast with currently available methods for general ERG models.  We propose to take advantage of this latter property, formulating our models in a fashion that facilitates computation for even very large, sparse dynamic graphs.  We also make use of available exponential family theory to derive a minimal set of assumptions that leads immediately to a lagged logistic form for the joint evolution of edge structure and vertex set.  This allows us to clarify what is being assumed in using such a model, thereby facilitating the assessment of its applicability in particular settings.  Moreover, placing this family within the general family of dynamic ERGMs allows it to be readily expanded by the incorporation of alternative dependence assumptions (although not without computational cost).  Key to our effort is the intuition that, in the dynamic case, \emph{the past history of the evolving network will account for much of the (marginal) dependence among edges}---thus, the assumption of conditional independence of edges in the present (given the past) may be a much more effective approximation for incremental snapshots of evolving networks than for typical cross-sectional and/or marginalized network data.  By leveraging this approximation, we can potentially account for many aspects of network evolution for systems whose size would prove prohibitive to more elaborate models.

The overall structure of the paper is as follows: we begin by describing the basic background and notation for our proposed modeling framework, following this with a derivation of the dynamic logistic regression family with vertex dynamics from the general family of dynamic ERGMs under specified independence assumptions.  We then consider computational issues, including scalability and fit assessment. Finally, we illustrate the use of this approach (and of associated adequacy diagnostics) via an application to the evolution of interpersonal communication of windsurfers on a beach in Southern California in the late summer of of 1986.  

\section{Notation and Core Concepts}
\label{bkg}

We begin by laying out the basic notation and statistical framework that underlies both the theoretical and methodological contributions of this work. This section first covers the necessary graph theoretic and matrix notation needed for defining the ERG models. We follow this with a brief review of core concepts from the ERGM literature that will be exploited in the subsequent sections of this paper.

\subsection{Graph Notation}

We here follow the common practice of representing structural concepts in a mixture of graph theoretic and statistical notation \citep[see, e.g.][]{wasserman.faust:1994a,butts08}. A \emph{graph} in mathematical language is a relational structure consisting of two elements: a set of \emph{vertices} or \emph{nodes} (here used interchangeably), and set of vertex pairs representing \emph{ties} or \emph{edges} (i.e., a ``relationship" between two vertices). Formally, this is often represented as $G=(V,E)$, where $V$ is the \emph{vertex set} and $E$ is the \emph{edge set}.  If $G$ is undirected, then edges consist of unordered vertex pairs, with edges consisting of ordered pairs in the directed case; our development applies in both circumstances, unless noted otherwise. 

Here we will represent the number of elements in a given set with the cardinality operator $| \cdot |$, such that $|V|$ and $|E|$ are the number of vertices and edges in $G$, respectively. The term for the number of vertices in a given graph in social network analysis is known as either \emph{order} or \emph{size} and is denoted $n=|V|$.  As noted below, we will be considering cases in which neither $E$ nor $V$ are fixed, but evolve stochastically through time.  Throughout, however, we will treat $n$ as finite with probability 1, and assume that the elements of $V$ are identifiable.

A common representation of graph, $G$, is that of the \emph{adjacency matrix} $Y,$ such that $Y= (y_{ij})_{1\leq i,j\leq n}$, where $y_{ij} = 1$ if $i$ sends a tie to $j$, or $0$ otherwise.  If $G$ is undirected then its adjacency matrix is by definition symmetric, i.e. $y_{ij}=y_{ji}$; if $G$ is directed then its adjacency matrix is not necessarily symmetric. It is common to assume that there are no self-ties (or \emph{loops}) and thus the diagonal is represented either as all zeros, ($y_{ii}=0$, or treated as missing, $y_{ii}=\mathrm{NA}$). This assumption is not necessary for the development that follows.

A necessary addition to this notation is that of an index for time, $t$, such that $Y$ becomes a $t$-indexed vector of adjacency matrices with $Y_{t}$ being a convenient shorthand for the adjacency matrix at time $t$, and $Y_{tij}$ an indicator for the state of $i,j$ edge at said time. We also apply this notation to graphs, such that $G_t=(V_t,E_t)$ denotes the state of $G$ at time $t$.  Our development assumes that $G$ is observed at a finite number of time points (i.e., we consider network evolution in discrete time).

\subsection{Random Graph Models and Exponential Family Form}

When modeling networks, it is helpful to represent their distributions via random graphs in exponential family form.  The explicit use of statistical exponential families to represent random graph models was introduced by \cite{holland81a}, with important extensions by \cite{frank86} and subsequent elaboration by \cite{wasserman96} and others.  Often misunderstood as a type of model \emph{per se}, the ERG (exponential-family random graph) formalism is in fact a \emph{framework} for representing distributions on graph sets, and is complete for distributions with countable support (i.e., one can always write such a distribution in ERG form, albeit not always parsimoniously).  The power of this framework lies in the extensive body of inferential, computational, and stochastic process theory (borrowed from the general theory of discrete exponential families) that can be brought to bear on models specified in its terms \citep[see, e.g.][]{barndorff78,brown86}; effectively, the ERG form constitutes a general ``language'' for expressing and working with random graph models.

Given a random graph $G$ on support $\mathcal{G}$, we may write its distribution in exponential family form as follows:
\begin{equation} \label{ergm}
Pr(G=g \: | \: s, \theta ) =\frac{ \exp \left(\theta^T s(g) \right) }{\sum_{g^{'} \in \mathcal{G}} \exp \left(\theta^T s(g^{'}) \right)} \mathbb{I}_\mathcal{G}(g),
\end{equation}
where $Pr(\cdot)$ is the standard probability measure, $\mathcal{G}$ is the support of $G$, $g$ is the realized graph, $s$ is the function of sufficient statistics, $\theta$ is a vector of parameters, and $I_\mathcal{G}$ is the indicator function (i.e. 1 if its argument is in the set-space of $\mathcal{G}$, 0 otherwise). 

While the extreme generality of this framework has made it attractive, model selection and parameter estimation are often difficult due to the normalizing factor in the denominator of Equation~\ref{ergm} (which is effectively incomputible except in special cases such as the the Bernoulli and dyad-multinomial random graph families \citep{holland81a}). The first applications of this family (stemming from Holland and Leinhardt's seminal 1981 paper) focused on these special cases.  \citet{frank86} introduced a more general estimation procedure based on cumulant methods, but this proved too unstable for general use; emphasis then switched to approximate inference using maximum pseudo-likelihood estimation \citep{besag1974}, as popularized in this application by \citet{strauss} and later \citet{wasserman96}.  Although maximum pseudo-likelihood estimation (MPLE) coincides with maximum likelihood estimation (MLE) in the limiting case of edgewise dependence, the former was found to be a poor approximation to the latter in many practical settings, thus leading to a consensus against its general use (see, e.g., \cite{besag00} and \cite{duijin07}).  The development of effective Markov chain Monte Carlo strategies for simulating draws from ERG models in the late 1990s \citep{anderson99a,snijders02} led to the current focus on MLE methods based either on first order method of moments (which coincides with MLE for this family) or on importance sampling \citep{geyer92}.  Algorithms for parameter estimation and model selection using these approaches are implemented in a number of software packages \citep[see, e.g.,][]{siena07,handcock.hunter.butts.goodreau.morris:2003a,PNet}, and empirical applications are increasingly common \citep[e.g.,][etc.]{goodreau09,Snijders20101,robins01}.

This tension between the capacity of the ERGM framework to represent computationally difficult models with substantial dependence and the need for models that can be deployed in practical settings has been a defining theme of research in this area.  In this paper, our concern is primarily with the latter problem: we seek families of models for network dynamics that are computationally tractable, and easily interpreted.  At the same time, however, we recognize the power and flexibility of the ERGM representation, particularly as a tool for embedding simple models within a much broader family (thus paving the way for subsequent expansion).  As such, we will draw heavily on the exponential family framework in our development, even when working with cases that can be represented in other ways (e.g., logistic regression).

\section{Modeling Network Dynamics with Logistic Regression}
\label{sec:dnlr}

Consider a discrete time series $\ldots,Y_0,Y_1\ldots$, where $Y_i \in \{0,1\}$.  One approach to modeling such a series is to posit that each $Y_i$ arises as a Bernoulli trial whose parameter, $\phi_i$, is the inverse logit of some given function of $Y_{i-1},Y_{i-2},\ldots$ (along, perhaps, with some vector of covariates $X_i$).  This model family is equivalent to logistic regression of $Y$ involving one or more ``lagged'' terms (i.e., functions of the prior values of $Y$), and is thus referred to as \emph{lagged logistic regression} (a natural analog of the Gaussian AR process \citep{brockwell.davis:bk:1991,shumway06}).  Models with lagged logistic form have been used for studying network dynamics, but the family as a whole has a higher level of generality than has been exploited in the social network literature.  In the development that follows, we review and extend the derivation of an analogous family of processes for dynamically evolving network data.  In keeping with the analogy, we refer here to the models associated with these processes as \emph{dynamic network logistic regression} or \emph{lagged network logistic regression} models.  Although this family lacks the full flexibility of the general ERGMs cited above, it has the advantage of being simple, scalable, and easily extensible to the case of \emph{network vital dynamics} (the ``birth'' and ``death'' of vertices).  These features make this model family a natural starting point for dynamic network modeling on large graph sequences.  Even where the family proves inadequate unto itself, its extensibility provides a natural path for incorporation of more complex forms of dependence. 

As noted, an important consideration in our development is \emph{scalability} to graphs with large vertex sets.  Recent innovations in data collection, as well as new forms of social interaction (e.g., online social networks) have greatly expanded the size of social networks available for study.  While this has been a boon to analysts, it has also posed significant challenges: the computational complexity of many basic network properties grows rapidly with the size of the vertex set, and the Monte Carlo procedures underlying conventional statistical procedures for network modeling require that such properties be evaluated large numbers (e.g., millions) of times.  These complexity problems are exacerbated in the dynamic case by the need to perform such computations for multiple temporal cross-sections. It is worth noting that computational power and algorithmic efficiency both continue to improve with time; however, at this current juncture, current implementations of general frameworks such as the actor-oriented models of \cite{snijders01} or the dynamic ERG models \cite{krackhardt07,krivitsky} are often impractical to apply to networks having even a few thousand nodes. Although scalability is a challenge for virtually all non-trivial network models, simplifying assumptions can often allow efficiency gains that permit the analysis of data that would otherwise be out of reach of statistical procedures.  We now turn to a consideration of one such set of assumptions, which jointly imply a general conditional logistic structure for networks with jointly evolving edge and vertex sets.

\subsection{The Core Dependence Structure}

In the conventional, cross-sectional case where $V$ is fixed, logistic models arise from the assumption that all edges are independent conditional on a fully-observed set of covariates \citep{wasserman.robins:ch:2005}.  Although potentially adequate in networks with very strong covariate effects \citep{butts:ch:2003}, such models are often poor approximations where covariate information is limited, or where complex interactive processes are the primary drivers of tie formation and dissolution \citep{goodreau09}.  Consider, however, the case of network ``panel'' data, in which an evolving network is measured at regular intervals during its evolution.  Here, too, simultaneity can be a problem, and specialized modeling schemes like those of \cite{snijders01}, \cite{krackhardt07} and \cite{krivitsky} have been proposed to capture this dependence.  If the intervals over which we measure the network are suitably fine, however, very little \emph{simultaneous} dependence is likely to occur: for many systems, much of what transpires over a short time interval can be treated as independent given the past history of interaction, and suitable covariates.  (Indeed, taking this logic to its infinitesimal extreme results in the relational event framework of \citet{butts08}, which exploits this property to measure the dynamics of event-based interaction in continuous time.)  Where this assumption is reasonable, it may be possible to approximate the process of network evolution by an inhomogeneous Bernoulli graph process in which edge states at future times depend upon the past history of the network, but not (conditionally) other edges at the same time point.  Such an approximation would allow one to leverage the substantial computational and interpretive advantages of the \emph{General Linear Model (GLM)} framework, while still capturing the critical mechanisms of network evolution.  

\begin{figure}[h]
    \centering
    \textbf{Dependence Diagram}\\

\includegraphics[width=.75\linewidth]{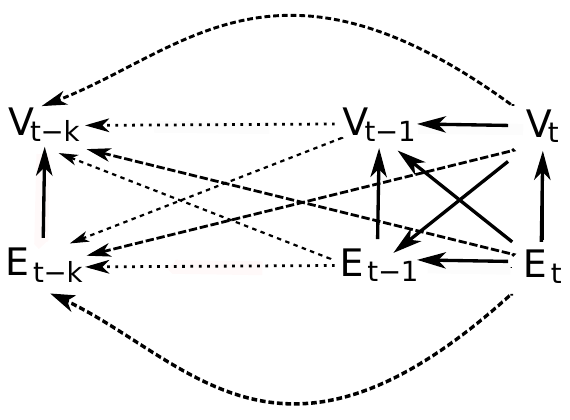}

    \caption{Representation of the dependence graph of the cross-sectional vertex and edge sets under the assumptions of Section \ref{sec:derlik}. $t$ represents time and $k$ represents the number of lags. }
    \label{fig:dep}
\end{figure}

The model family we propose is one that leverages \emph{potentially complex dependence on the past} together with \emph{conditional independence in the present} to flexibly capture network evolution in a way that nonetheless reduces to lagged logistic regression.  Specifically, we derive our model family from the core assumption that $E_{t+1}$ depends only on $V_{t+1}$ and $(E_t,V_t),\dots,(E_{t-k},V_{t-k})$, and $V_{t+1}$ depends only on $(E_t,V_t),\dots,(E_{t-k},V_{t-k})$, together with any exogenous covariates (see Figure \ref{fig:dep}). Intuitively, this can be thought of as specifying that today's vertices are determined by the past network structure (out to some limit, $k$), and that today's edges are determined by both this past structure \emph{and} today's vertices.  One of the effects of this framework is that it allows uncertainty in network composition to be considered when making predictions.  As we shall see, explicitly considering this aspect of network structure (which has been largely overlooked in prior research) leads to a very different view of network dynamics in contexts for which vertex entry and exit are possible.

\begin{center}
[ Figure \ref{fig:dep} ]
\end{center}

Although the aforementioned model family treats edges as conditionally independent within time steps, they may depend upon past time steps via arbitrary functions of previous graph realizations (up to some finite order, $k$).  We call such functions of previous network states \emph{lag terms} (in analogy with time series models), with the \emph{order} of a lag term corresponding to the temporal difference between the earliest cross-section employed by the term and the current cross-section.  (Thus, a first order term involves only the previous time step, the second involves at most the second, etc.)  In general, our framework allows for arbitrary choice of $k$ (and thus dependence over arbitrarily long lags).

\subsection{Deriving the Likelihood}
\label{sec:derlik}

To obtain the dynamic logistic network regression representation for our process, we break down the derivation into two distinct parts. First we define the necessary assumptions for the likelihood of the relational structure of the graph given the vertex set, and next we define the necessary assumptions to derive the fully logistic structure for modeling both the vertices and the edges as a lagged logistic regression model. Note that unlike the preceding sections where we employed the edge set notation ($E$), we now apply the adjacency matrix notation ($Y$) in the following section for greater flexibility in handling edge set decomposition. 

We start by relaxing the temporal Markov and fixed vertex set assumptions of \citet{hanneke07}, replacing them with weaker versions.  We then impose some conditional edge and vertex independence assumptions, and lastly we make some homogeneity assumptions.  We formally specify these assumptions in Section \ref{part1} and \ref{part2}, combining them to derive the likelihood of the dynamic network logistic regression model family. 

This structure allows us two distinctive advantages over \citet{hanneke07} and others. The first advantage is that unlike \citet{hanneke07}, we do not require the vertex set to be fixed and thus the number and identity of vertices may change with time (an important factor when modeling emergent networks e.g. as arising following disasters, in naturally occurring groups, etc.). The second important distinction is that we explicitly develop the dependence conditions needed for inhomogeneous Bernoulli structure, in comparison to \cite{hanneke07} whose computational examples implicitly assume Bernoulli structure but who do not elaborate the associated theoretical assumptions.  This development facilitates the expansion of the present model family by relaxation of conditional independence, where necessary.

\subsubsection{Part 1: Edges Given the Vertex Set}
\label{part1}

We consider first the evolution of edges, given the vertices present in the network. Given a graph $G_i \sim (Y_i,V_i) = Z_i$ and covariate set $X_t$ (noting that $X$ may contain covariate information from prior time points) with $i\in 1,\ldots,t$, we formally specify our assumptions below: (\ref{asone}) states that the state of the network at any given time point depends only on the states of the networks over some previous $k$ time points (the relaxed temporal Markov assumption); (\ref{astwo}) asserts conditional independence of edges in the same time slice, given past history and covariates; and (\ref{asthree}) and (\ref{asfour}) assert that the stochastic process generating the network is temporally homogeneous (given the covariates). 

\begin{enumerate}[(i)]
\item For some specified $k\ge 0$, $Z_i\: | \: \left\{Z_{i-1},\dots,Z_{i-k},\: X_t \right\}$ is independent of $Z_{i-k-\delta}$ for all $\delta>0$. \label{asone}
\item $Y_{ijk}$ is independent $Y_{igh}$ given $\left\{V_i,Z_{i-1},\dots,Z_{i-k}, \: X_t \right\}$ for all $j,k \neq g,h$. \label{astwo}
\item  Let $f_Y$ be the conditional pmf of $Y_i$ (i.e., an arbitrary time slice of $Y$).  For any realizable $y,y_1,y_2,\ldots,y_k$, $v,v_1,v_2,\ldots,v_k$ then, for all $i,j \in 1,\ldots,t$: \label{asthree}
\begin{eqnarray*}
 \lefteqn{f_Y(Y_i=y \: | \: V_i=v,Z_{i-1}=z_1,\dots,Z_{i-k}=z_k, \: X_t=x_t) =} \\
& f_Y(Y_j=y\: | \: V_j=v,Z_{j-1} = z_1,\dots,Z_{j-k}=z_k, \: X_t=x_t). 
 \end{eqnarray*}
 \item Let $f_V$ be the conditional pmf of $V_i$ (i.e., an arbitrary time slice of $V$).  For any realizable $y_1,y_2,\ldots,y_k$, $v,v_1,v_2,\ldots,v_k$ then, for all $i,j \in 1,\ldots,t$:  \label{asfour}
 \begin{eqnarray*}
  \lefteqn{f_V(V_i=v \: | \: Z_{i-1}=z_1,\dots,Z_{i-k}=z_k, \: X_t=x_t) =} \\
  &  f_V(V_j=v \: | \: Z_{j-1} = z_1,\dots,Z_{j-k}=z_k, \: X_t=x_t).
\end{eqnarray*}
\end{enumerate}

\noindent
From these assumptions, we can derive the joint likelihood of the network time series.  We begin by applying assumption (\ref{asone}), which allows us to decompose the joint likelihood of the time series as a product of conditional distributions:
\begin{eqnarray}
 & Pr((Y,V)=(y,v) \: | \: X_t) = \prod_{i=k}^t Pr(Z_i= Z_i\: | \: Z_{i-1},\dots,Z_{i-k}, \: X_t) \nonumber
\end{eqnarray}      

\noindent
Applying assumption (\ref{astwo}) we can further decompose the joint likelihood into vertex and adjacency components, the latter written as products over individual edge variables:

\begin{eqnarray}
\lefteqn{= \prod_{i=k}^t Pr(Y_i=y_i \: | \: V_i,Z_{i-1},\dots,Z_{i-k}, \: X_t)  \times Pr(V_i=v_i \: | \: Z_{i-1},\dots,Z_{i-k}, \: X_t)}  \nonumber \\
\lefteqn{ = \prod_{i=k}^t Pr(V_i=v_i \:| \: Z_{i-1}\dots,Z_{i-k},X_t) \times} \\
& \: \: \; \: \: \: \prod_{i=k}^t \prod_{(g,h)\in V_i^2} Pr(Y_{igh}=y_{igh} \: | \: V_i,Z_{i-1},\dots,Z_{i-k}, \: X_t) \nonumber
\end{eqnarray}

\noindent
Homogeneity assumptions (\ref{asthree}) and (\ref{asfour}) allow the above to be written in terms of the pmfs $f_V$ and $f_Y$:

\begin{eqnarray}
&= \prod_{i=k}^t f_V(v_i \: | \: Z_{i-1},\dots,Z_{i-k},X) \times \prod_{i=k}^t \prod_{(g,h) \in V_i^2} f_Y(y_{igh} \: | \: V_i,Z_{i-1},\dots ,Z_{i-k}, \: X_t) \nonumber
\end{eqnarray}

\noindent
which by the completeness of the exponential family representation for a binary variable leads us to

\begin{eqnarray}
 \lefteqn{= \prod_{i=k}^t [f_V(v_i \: | \: Z_{i-1},\dots,Z_{i-k}, \: X_t)} \label{like1} \\
 &\times \prod_{i=k}^t \prod_{(g,h) \in V_i^2} \textrm{logit}^{-1}(u(y_{igh},V_i,Z_{i-1},\dots,Z_{i-k}, \: X_t)). \nonumber
\end{eqnarray}

\noindent
Thus, each adjacency snapshot is conditionally a logistic network model, and is separable from the likelihood of $V$.  \\

\subsubsection{Part 2: Vital Dynamics}
\label{part2}

There exist few inferential models in the social network literature which model the vital dynamics of a social network; however, vital dynamics can greatly influence the nature and characteristics of a given social network.  We propose using the aforementioned dynamic logistic regression as a reasonable starting point. As with edge dynamics, logistic structure for vertex entry (``birth") and exit (``death") arises naturally given a series of simplifying conditional independence assumptions.\footnote{Note that we do not require that vertices can enter or exit only once, although adding such an assumption may be appropriate in some settings.}

In order to model vital dynamics in a practical fashion, we propose the following additional simplifying assumptions. We begin with (\ref{asseven}), which simply states that there exists a finite set that contains all vertices at risk of entering the network over the entire time period $1,...,t$. Next, we make  another conditional independence assumption (\ref{asfive}) such that vertex set at time $V_t$ is conditionally independent of network realizations prior to a fixed point in the past. We then assume  (\ref{assix}) that the indicator of vertex  $g$ is conditionally independent of the indicator of vertex $h$, $h\neq g$, (i.e., whether vertex $g$ is present or not is conditionally independent of $h$) given the edges set at time $t$, the past realizations of the edge and vertex set, and exogenous covariates. Lastly, we make a homogeneity assumption (\ref{asseight}) that parallels that of the edge case.

\begin{enumerate}[(i)]
\setcounter{enumi}{4}
\item  There exists some finite set $V_\mathrm{max}$ such that $V_i\subseteq V_{max}$ for all $i\in 1,\ldots,t$. \label{asseven}
\item $V_i$ is independent of $Z_{i-k-\delta}$ given $Z_{i-1},\dots,Z_{i-k}, X_t$ for all $\delta>0$. \label{asfive}
\item $\mathbb{I}(g \in V_i)$ is independent of $\mathbb{I}(h \in V_i)$ given $Z_{i-1},\dots,Z_{i-k}, X_t$  for all $g \neq h$. \label{assix}
\item Let $f_{V,i}$ be the conditional pmf of inclusion for some vertex $i$ in some $V_j$.  Then, given any realizable $v_1,v_2,\ldots,v_k$ then, for all $i \in 1,\ldots,t$ and all $g,h \in V_\mathrm{max}$,\label{asseight}
\begin{eqnarray}
 \lefteqn{f_{V_g} \left(\mathbb{I}(g \in V_i) =1 \: | \; Z_{i-1} = Z_{i-1},\ldots, Z_{i-k} = Z_{i-k},  X_t=x_t \right)} \\
 &=f_{V_h}\left(\mathbb{I}(h \in V_i) \: | \: Z_{i-1} =Z_{i-1},\ldots, Z_{i-k}=Z_{i-k},X_t=x_t \right) \nonumber
\end{eqnarray}
 \end{enumerate}

\noindent
With assumptions (\ref{asseven}), (\ref{asfive}), (\ref{assix}), and (\ref{asseight}) and the exponential family argument applied earlier, we may rewrite the left hand side of equation \ref{like1}: 

\begin{eqnarray}
\lefteqn{f_V(V_i \: | \:V_{i-1},\dots,V_{i-k},X_t) = }\nonumber \\
&& \prod_{i=k}^t  f_V(\mathbb{I}(g \in V_i \: | \: \forall \: g \in V_{max} )\: | \: V_{i-1},\dots,V_{i-k}, \: X_t) \label{like2}  \\
 &&= \prod_{i=k}^t \textrm{logit}^{-1}(w(\mathbb{I}(g \in V_i \: | \: \forall \: g \in V_{max}),V_{i-1},\dots,V_{i-k},X_t) \nonumber  
 \end{eqnarray}

\noindent
Thus, with these additional constraints we acquire a \emph{dual-logistic structure}. We may thus summarize the likelihood of the vertex portion of the model and the edge portion of the model in seperable terms. The vertex likelihood is given by

\begin{equation}
Pr(V_t\:|\:Z_{t-1},\dots,Z_{t-k}) = \prod_{i=1}^n \textrm{logit}^{-1} \left(w(\mathbb{I}(v_i \in V_t),Z_{i-1},\dots,Z_{i-k})\right) 
\label{eq:vert}
\end{equation}

\noindent
and the edge likelihood by

\begin{equation}
Pr(Y_t\:|\: V_t, Z_{t-1},\dots,Z_{t-k}) = \prod_{(i,j)\in V_t \times V_t}^n \textrm{logit}^{-1} \left(u(Y_{tij},V_t,Z_{i-1},\dots,Z_{i-k}) \right), \label{eq:edge}
\end{equation}

\noindent
with the joint likelihood being the product of the two.  A useful computational side effect of this is that we may use a single logistic routine to fit the entire model, using the augmented vector of the adjacency matrix and the temporal vertex indicator set (Equation~\ref{eq:vert} and \ref{eq:edge}). 

The above provides a fairly flexible and highly tractable framework for modeling joint edge/vertex dynamics, for the case in which the risk set of potentially appearing vertices is known (or can be approximated as such).  In some cases, this risk set may be well-approximated by the set of all vertices ever appearing in the network (e.g., that the chance of a vertex being effectively at risk and never actually appearing is small).  In other cases, it may be desirable to consider a larger population of potential actors.  (We assume at present that this set is bounded, although extensions using dirichlet processes \citep{ferguson73} or the like could be employed to generalize this framework to the unbounded case.)  For inferential purposes, estimation for parameters of both vital dynamics and edge dynamics are performed within the same logistic regression, and are fully separable.  In the case of simulation, however, the dependence structure illustrated in Figure~\ref{fig:dep} requires alternately sampling vertices and (conditionally) edges on those vertices.  As this suggests, both edge and vertex submodels can interact in complex ways to create network structure, even where these models are inferentially distinct.  An example of this interaction is shown in Section~\ref{sec:application}.

\section{Practical Considerations: Scalability, Estimation, and Model Adequacy Assessment}
\label{}

When implementing and evaluating models of the type discussed here, there are several important practical considerations to be considered.  First is the issue of scalable implementation.  One advantage of the logistic framework is that there is a large body of work in computer science and machine learning regarding inference for logistic regression in large, sparse matrix settings, that can be utilized when fitting logistic models to large dynamic networks.  Second, the issue of parameter and variance estimates is of important concern, particularly to social scientists who employ coefficients to estimate the strength of putative tie formation mechanisms (as opposed, e.g., to ``black box'' forecasting). Third, some method of model evaluation is necessary so as to assure the analyst that the model captures the important macro-level characteristics of the graph which inform his or her theory.

The following three sections represent an integrated discussion of these issues. Following this section an application of these methods is demonstrated on a dynamic interpersonal communication network.

\subsection{Scalability}
\label{}

As noted earlier, logistic regression is a popular and well-established technique for statistical analysis \citep{mccullagh99}. Standard optimization techniques may be applied to logistic regression for quite large data sets with current technology (e.g. in Section \ref{sec:application} we employ a gradient based optimization technique on the full likelihood to an evolving network of 95 actors, and have had success with networks larger by an order of magnitude or more). 

The scalability of logistic regression has been of particular interest in the machine learning literature, as the approach is used on a wide range of problems such as as neural networks and binary classification \citep{devroye96}. The computer science community has thus spent significant amounts of time and energy in developing scalable logistic regression algorithms \citep{komarek04,komarek03,mccullagh99,lin08}. The current literature in machine learning focuses on four core methods: iterative-scaling \citep{darroch72,pietra97,goodman02,jin03}, nonlinear conjugate gradient \citep{vetterlin92}, limited memory quasi-Newton \cite[also known as L-BFGS methods,][]{liu89,benson01}, and truncated-Netwon \citep{komarek05}. \cite{malouf02} found that the limited memory quasi-Newton methods were the most efficient in a series of computational trials. Recently, \cite{lin08} have proposed and implemented \citep{fan08} a trust-region Newton Method for large-scale logistic regression based on the optimization technique of \cite{lin99}. \cite{lin08} successfully apply their method to data sets with hundreds of millions of observations and millions of covariates. Each of these methods typically involve clever ways of managing the linear algebra and derivative problems encountered in modern optimization problems. 

All of these methods are potentially applicable to the problem of dynamic network logistic regression. The richness of this literature and the constant growth in optimization of large-scale data problems allow the methods discussed in this paper to be applied to increasingly large data sets (large in time, vertex size or both).  While not all network time series require such methods, the latter's availability makes this approach particularly useful for cases in which more general models would prove computationally infeasible.

\subsection{Estimation: Bayesian Analysis and Bias Reduction}

In conducting likelihood-based inference via Equations~\ref{eq:vert} and \ref{eq:edge}, both frequentist (e.g., maximum likelihood) and Bayesian approaches from the standard literature may be employed.  In test cases (like that of Section~\ref{subsec:data}) we have obtained similar results from both standard maximum likelihood (ML) estimates and Bayesian posterior mode estimates with weakly informative Student's $t$ priors (in Section~\ref{subsec:data}, a $t$ prior centered at 0 with a scale parameter of 2.5 and one degree of freedom, i.e. a Cauchy distribution, is employed). In conventional logistic regression settings, \cite{gelman08} recommend a $t$ prior distribution as the default choice for routine use. They argue that it has the advantage of always yielding a well-defined posterior estimate, and automatically applying more shrinkage to higher-order interactions. \cite{gelman08} derive a modified EM algorithm to produce the parameter and error estimates.  The analyst may interpret the resulting estimator in either frequentist or Bayesian terms. From a Bayesian point of view, the estimator in this case is the mode of the posterior distribution where all model parameters are viewed a priori as having a multivariate $t$ distribution, an estimator which is optimal under 0/1 loss.  Within a frequentist framework, the use of a ``prior'' structure may be thought of as a bias reduction technique.  As past work on related models has suggested that estimates of uncertainty are better-behaved under this alternate procedure than estimates obtained from the Hessian of the deviance matrix, we recommend the use of the former in typical settings. 

\subsection{Model Adequacy Assessment and Simulation Analysis}
\label{subsec:gof}

Model selection and assessment is a common problem in all fields employing mathematical and statistical models. In this paper we begin by distinguishing between model selection and model assessment. For the former problem, we recommend that the analyst start with the standard model selection techniques based on penalized log-likelihood approaches such as the Bayesian information criterion (BIC) \citep[][]{schwarz78} or Akaike information criterion (AIC) \citep[][]{akaike74} to deciding which model performs best within a collection of proposed models. This procedure follows standard statistical practice, and is reasonable well-developed; for further details see \citet{brockwell.davis:bk:1991,gelman03}.  Given that one has identified the best overall candidate model, we then recommend performing simulation-based assessments of model adequacy to verify that the candidate captures the relevant properties of the original data; the approach to adequacy testing suggested here is an adaptation and extension of those applied in the computational Bayesian literature \citep{gelman03} and the model assessment methods for cross-sectional network data \citep{hunter08}. 

Modern network analysis often applies simulation-based methods for analysis, prediction, exploration or model diagnostics. Simulation is typically used in these cases because few network models lend themselves to analytical treatment.  In this paper we employ simulation methods in order to ascertain the model performance on a series of theoretically motivated network metrics (i.e., model adequacy assessment).  

In the machine learning literature there are a number of different approaches to prediction, one of which is known as the 50-percent classifier rule \citep{devroye96}. The 50-percent classifier rule is a threshold model (0.50) where it is assumed that an event occurs if the predicted probability of the event occurs at over a half. This predictive model may be applied to an in homogeneous  Bernoulli structure in a straightforward manner: apply this threshold to each predicted probability, in this case first to the vertex set and then to the resulting edge set predictions. It is quite natural to generalize this basic approach through simulation (i.e., apply a Bernoulli process to each predicted probability and use a computer to generate $n$ predictions (0 or 1) from each given probability).  We refer to this technique as an \emph{inhomogeneous Bernoulli classifier}. This method allows for a full assessment of predictive uncertainty of the inferred model under the assumed conditions. 

The algorithm we employ is as follows (Algorithm~\ref{alg:bc}): for each time point ($t$) we predict $n$ observations one-step ahead (i.e., we predict time point $t$ from time $t-1$) by applying the aforementioned inhomogeneous Bernoulli classifier, where first we predict the vertex set (e.g., the vertices that we project to occur at time $t$), and then from the vertex set we predict the edge structure. Then we save a set of well chosen Graph Level Indices (GLI) \citep{wasserman.faust:1994a, anderson99} (so that we are not required to store $n$ graphs a $t$ time points, which could become computationally impractical in many of the desired cases for this model).  

\begin{center}
[Algorithm~\ref{alg:bc}] 
\end{center}

The reason for concentrating on GLI distributions is twofold: first, it is often difficult or impractical to visually inspect thousands of simulated networks, nor are these easy to compare statistically in simple and practical terms without the use of descriptive indices.  Second, it is typically the case that the analyst is not concerned with the occurrence of a single edge or vertex, but rather with the overall macro-level properties of the network (e.g. mean degree, triad census, centrality measures, connectedness measures, and so forth).  Examination of a limited set of index distributions accomplishes the latter goal, while avoiding the former difficulty.

After we perform the simulation procedure, we say that the proposed model ``adequately'' captures a given feature of the observed network at a specified level of precision $\alpha$ if the associated GLI value falls within the central $\alpha$-coverage simulation interval for the model in question.  The optimal case is naturally one in which the simulated GLI distribution is centered on the observed value, with little variation; for a simple model of a complex system, however, we may employ a looser criterion (e.g., coverage by the 95\% simulation interval for a certain fraction of time steps).  Selection of both GLIs to study and adequacy criteria are necessarily dependent upon substantive considerations (including the use to which the model is to be put).  For example, if one's central theoretical concern is the explanation of transitivity in an evolving network, then ensuring that this index is well-accounted for by the model (in the sense of reliably included in simulation intervals with $\alpha\le 0.95$) would be critical.  In the same context, one might be less concerned with capturing, say, mean degree, but may nevertheless show concern if such a basic property were not covered by wide (say, 99\%) simulation intervals in a significant fraction of time points.  For an extensive example of this procedure see Section~\ref{subsec:gof2}.

\section{Sample Application: Going to the Beach}
\label{sec:application}

To illustrate the application of the dynamic network logistic regression approach, we employ the methods discussed in this paper to the analysis of a classic network data set. This data involves a dynamically evolving network of interpersonal communication among individuals congregating on a beach in Southern California observed over a one-month period \citep{freeman88,freeman92}. Interpersonal communication in small groups is a well studied subfield in social psychology and social network analysis \citep{festinger51}. The importance of studying interpersonal communication networks dynamically was originally pioneered by \cite{nordlie58} and \cite{newcomb61}; here, we show how the dynamic logistic family allows us to flexibly model the evolving network, with particular emphasis on the interplay between tie structure and vertex set dynamics. 

\subsection{Data}
\label{subsec:data}

The data analyzed in the following sections was originally collected and analyzed in aggregate by \cite{freeman88} and has since been used in a number of influential articles \cite[see][etc.]{benjamin09,hummon03,zegglink96}. While this network is typically analyzed in aggregate, it was originally collected as a dynamically evolving network (where the vertex set is composed of windsurfers and the edge set is composed of interpersonal communication). The network was collected daily (sampled at two time points each day) for 31 days (August  28, 1986 to September 27, 1986).\footnote{Unfortunately, one day (September 21st) is missing due to a race on a different beach, which precluded data collection.  Thus, complete data is available for 30 days during the observation period.} 

Individuals were tracked with a unique ID, and were divided by Freeman et al. into those we will here call \emph{``regulars''} ($N=54$) -- frequent attendees who were well-integrated into the social life of the beach community -- and \emph{``irregulars''} ($N=41$) on ethnographic grounds.  The former category was further broken down by the researchers into two groups, \emph{Group 1} ($N=22$) and \emph{Group 2} ($N=21$), with 11 individuals not classified as belonging to either Group 1 or Group 2. Altogether, the union of vertex sets ($V_{\max}$) consists of 95 individuals. On any given day during the observation period, the number of windsurfers appearing on the beach ranged from 3 to 37, with the number of communication ties per day ranging from 0 to 96. 

These basic characteristics will be used in the illustrative analysis that follows, which centers on the question of what drives the evolution of interpersonal communication in this open, uncontrolled setting.

\subsection{Mechanisms of Dynamic Interpersonal Communication} 
\label{subsec:mech}

A number of distinctive mechanisms may influence whether a windsurfer engages another windsurfer at any given time; however, two windsurfers clearly cannot interact if both do not appear simultaneously on the beach, and thus the first influences to be considered are those affecting the vertex set. For this illustrative analysis, we propose four basic mechanisms for the propensity of an individual to appear on a given day:  (1) a regularity effect; (2) an inertial network effect (e.g., the lag term); (3) a three-cycle effect (because this graph is symmetric this is equivalent to a triadic term); and (4) seasonal effects (e.g., day of week).  An intuitive summary of each mechanism follows. 

Of the four mechanisms we consider as drivers of vertex set dynamics, the first is \emph{regularity}, the notion that an individual is more likely to appear on any given day if he or she is one of the individuals who is classified (on ethnographic grounds) as belonging to the category of ``regulars'' who form the core of the beach community.  This recognizes the fact (known from the observational accounts) that there is heterogeneity among the windsurfers, with certain individuals being much more active than others.

The second posited mechanism is one of \emph{persistence} or \emph{inertia} -- i.e., if an individual is active today, he or she is more likely to be active or have tomorrow.  This is sometimes known in the social network literature as ``behavioral inertia" and has been seen both empirically and experimentally in varied social network contexts \citep{corten10}.

The third mechanism is a \emph{triangle effect}, where the number of three-cycles in which an individual is embedded at point $t-k$ influences the likelihood of whether an individual will appear on day $t$. This may be thought of as capturing the effect of social participation, with the intuition that persons embedded in dense social groups (e.g., cliques) are more likely to have their attendance reinforced, and thus to return to the beach.

The fourth mechanism is \emph{seasonality}, i.e. the tendency for activity to show systematic variation over daily or weekly cycles.  Cyclic phenomena are common in human systems, as has long been recognized in the time series literature \citep{shumway06}. Common seasonal effects in behavioral data include daily and hourly effects (e.g., differences between weekday and weekends, or midnight versus midday).  Networks are no exception to this rule, as evidenced by Baker's (\citeyear{baker:ajs:1984}) observation of daily variation in structure and activity within trading networks in a national securities market, and Butts and Cross's (\citeyear{butts09b}) finding that the volatility of evolving blog citation networks changes with time of day, day of week, and external events (in that particular case, phases of the 2004 US electoral cycle).  In the present case, a parallel phenomenon may occur through weekly cycles in the frequency of attendance at the beach (a reasonable expectation given the institutional context of work and leisure time for the study population during this period). 

Once the vertex set arises, the influence of a new set of interpersonal communication mechanisms becomes relevant.  Of the many potential mechanisms that could govern interpersonal communication in the study population, we here explore six: (1) regularity of beach use and other assortative mixing effects; (2) individual propensity effects for regularly occurring individuals; (3) contagious participation; (4) inertial network effects (e.g., the lag term); (5) embeddedness;  and (6) seasonal effects.  As with the vertex model, we briefly consider each of these in turn.

The first mechanism is \emph{assortative mixing} between those identified as regular beach goers and those who were classified as irregulars. In the social network literature, effects of a priori group partitioning on tie formation are often referred to as \emph{mixing} effects. \cite{mcpherson01} review extensive evidence that individuals cluster on homophilous grounds, and thus we might expect that those more deeply embedded in the milieu of the beach environment (the ``regulars'') will be more likely to talk with others of the same ilk (and, likewise, that outsiders will be more likely to interact with other outsiders).  Furthermore, among the regulars, those identified as belonging to the same core groups by the ethnographic observers are conjectured to mix at higher rates, ceteris paribus, than others.  

The second mechanism consists of \emph{individual-level heterogeneity} in the propensity of regular attendees to engage in communication with others. We might expect that idiosyncratic shyness or gregariousness of regularly occurring individuals may influence the amount of activity on a given day. Similar to the argument applied for the first mechanism we might expect the basic propensity of a regular attendee to engage or not engage other beach members to be highly influential on the amount of activity on any given day.

The third mechanism is \emph{contagious participation}, based on the notion that high levels of beach-going activity at the group level are likely to translate into high levels of other activity (including communication).  Thus, we take the number of persons present itself as a predictor of the propensity of individuals to communicate with others on the beach.

The fourth mechanism is \emph{persistence} or \emph{inertia} -- i.e., if an individual is active or has a relation today, he or she is more likely to be active or have a relationship tomorrow. 

The fifth mechanism is \emph{embeddedness} \citep[see][]{granovetter85}. A dyadic relationship which is embedded within a broader communicative context -- e.g., in which there persons in question are linked by numerous past chains of communication -- is likely stronger and more likely to persist at a later time point than one lacking such a context.  We measure embeddedness by the number of $k$-cycles within which a given relation is embedded.  

The sixth mechanism is \emph{seasonality}, here in the propensity to form ties rather than the tendency to appear at the beach.  This might arise for a number of reasons, e.g., systematic variation in the sorts of people who go on weekdays versus weekends, differences in activities pursued during weekday versus weekend excursions, and so forth. 

Each of the proposed mechanisms for both vertex formation and edge creation may or may not be important to the network structure, which brings up the necessary process of model selection and model adequacy assessment. In the following sections we employ first a penalized deviation model assessment to select the best fitting model. We then employ a series of simulation-based model adequacy checks as discussed in Section~\ref{subsec:gof} to assess the extent to which the selected model does or does not capture important features of the evolving network.

\subsection{Model Selection and Adequacy}
\label{subsec:gof2}

\subsubsection{Parameterization}

To implement our model, the impact of each of the mechanisms in Section~\ref{subsec:mech} is operationalized as a \emph{weight} or \emph{parameter} in the dynamic logistic regression framework. The first step in the model-building process is to select the vertex mechanisms, which are highly influential in this context because the vertex portion of the model predicts ``who shows up to the party" (so to speak), and thus who is eligible to interact at a given time point. The importance of ``who shows up" will greatly depend on the context and actor-specific covariates in a given dynamic network. For the beach data (as we will see) the most important attribute that an individual carries with him or her is whether or not he or she is a regular beach attendee (and which group within the regular attendees he or she is). If more information about these windsurfers had been collected we might, for example, expect there to be a gender effect and/or a ``couple" effect. It should also be noted, however, that individuals carry more with them than their exogenous covariates: insofar as an individual's interaction history affects his or her probability of communicating with others, he or she is less substitutable with peers having different histories of interaction.  Thus, correct prediction of individual attendance can be important even in settings for which exogenous covariates are limited (or altogether absent).

In addition to specifying putative mechanisms, our vertex model requires specification of the risk set ($V_{\max}$), i.e. the set of persons effectively at risk for showing up on a given day.  Here, we treat all individuals observed at any time during the data collection window as our risk set, lacking other information on potential attendance.  While this is obviously a simplification, we view the total set of all persons appearing over an entire month as a reasonable proxy for the unobserved collection of persons with a non-small chance of appearing on any given day.

As with other exponential family models, we capture the effects of putative mechanisms by statistics that (together with their associated parameters) determine the probability that an edge or vertex will appear at a given point in time. In describing these statistics, we employ the following notation.  Within this section, $t$, $i$, and $j$ jointly index the adjacency structure, so that e.g. $Y_{tij}$ represents the edge between the $i$th and $j$th vertices of $V_{\max}$ at time $t$.  Time itself is indexed in integer increments from $1,\ldots,T$, e.g., $T=31$ for the beach network.  We will frequently use $k$ to represent lags, e.g. with $Y_{t-k}$ representing the state of the edge set at time $t-k$.  The vertex and edge model statistics themselves follow the basic form of Section~\ref{sec:derlik}, with $w_{tp}^{\cdot}(V,Y,X)$ being a generic function for a statistic at vertex $p$, and $u_{tij}^{\cdot}(V,Y,X)$ being a generic function for a statistic at edge $ij$. $X$ represents the relevant covariates for a vertex or edge (i.e., $X_p^{\cdot}$ is a dichotomous variable for whether $v_p$ is a regular ($r$) or irregular ($\delta$); $X_{ij}^{\cdot}$ is a dichotomous variable for whether edge $ij$ is regular ($r$), irregular ($\delta$), or regular to irregular and visa versa ($\phi$); and $X_{tp}^d$ is the day (Monday,\dots, Sunday) at time $t$ for vertex $v_i$ and $X_{tij}^d$ is the day at time $t$ for edge $ij$. For simplicity in notation, we also define two measures: (1) $\tau_{tp} =$ the count of triangles within which $v_p$ is embedded at time $t$; and (2) $\zeta_{tij}^\epsilon=$ the count of $\epsilon$-length cycles within which edge $ij$ is embedded. 

To implement our covariate effects, we employ a series of dummy variables for whether an individual is in the regular category or in Group 1 category ($w_{tp}^{r}(V,Y,X) = X_{p}^{r}$)\footnote{We also tested Group 2, but this group was not particularly influential in either the vertex predication or interaction of individuals.}. For the inertial mechanism, we employ a single lag term with the basic interpretation that if this weight is positive than an individual is more likely appear on a given day if he or she was at the beach the day before ($w_{tp}^{l}(V,Y,X)= \mathbb{I}\{v_{p} \in V_{t-k}\}$, i.e. one if the focal actor was present at time $t-k$ and zero otherwise). For the triangle effect we employ three-cycle lag statistic  with the interpretation that a vertex is more likely to appear on a given day if he or she was embedded in a triangle relation the day before ($w_{tp}^{\Delta}(V,Y,X) =\tau_{t-k,p}$, i.e. the number of 3-cliques in which the focal actor participated at time $t-k$). We employ a dummy variable for each day of the week, thus allowing for higher or lower likelihood of every individual appearing on a given day of the week ($w_{tp}^s(V,Y,X) = (\mathbb{I}\{X_{tp}^d = Tuesday\},\dots, \mathbb{I}\{X_{tp}^d = Sunday\})$, with Monday as the reference category).  

As with the vertex model, operationalization of the edge model is performed by mapping the putative edge formation mechanisms onto a set of sufficient statistics.  Per the previous discussion, the mechanism of assortative mixing between regulars and irregulars is implemented as three statistics ($u_{tij}^r(V,Y,X) =X_{ij}^r$, $u_{tij}^{\delta}(V,Y,X) = X_{ij}^{\delta}$, and $u_{tij}^{\phi}(V,Y,X) =X_{ij}^{\phi}$) where the first represents the baseline effect of regular to regular interaction, the second represents irregular to irregular interaction and the last represents regular to irregular (and visa versa) interaction (noting that this term stands in place of the standard intercept term). The mechanism of individual-level heterogeneity is implemented as a dummy variable for each of the Group 1 members ($u_{tij}^h = \mathbb{I}\{ v_i \textrm{ or } v_j \in \textrm{ Group 1 } \}$)\footnote{We also tested all regulars and Group 1 and Group 2 individuals, and just Group 2 individuals, but found that Group 1 individuals were the set of most influential actors in this case.}. The mechanism of contagious participation is implemented as a density effect ($u_{tij}^{c}(V,Y,X) = \log(|V_t|)$) which changes dynamically based on the log of the number of individuals at the beach on the given day of interest (exploiting the fact that, because each day's edge realization is conditioned on that day's vertex set, properties of the latter can be used to predict the former).  The mechanism of inertia is implemented as a single lag term ($u_{tij}^{l}(V,Y,X) = Y_{t-k,ij}$).  The embeddedness effect is implemented as the log of the dyadic count of the number of cycles (up to 9) of the lagged network ($u_{tij}^{e}(V,Y,X) = \log(\zeta_{t-k,ij}^\epsilon+1)$), with the interpretation that a dyadic interaction is more or less likely if the edge existed yesterday and was in more or fewer cycles (depending on the sign of the weight). The last mechanism, seasonality, is again implemented as a series of dummy variables for each day of the week, with Monday as the reference category ($u_{tij}^{s}(V,Y,X) = (\mathbb{I}\{X_{tij}^d = Tuesday\},\dots,\mathbb{I}\{X_{tij}^d = Tuesday\})$).

\subsubsection{Model Fit}

Each mechanism proposed in Section~\ref{subsec:mech} may or may not influence whether a windsurfer arrives on a given day and/or whether or not he or she interacts with another windsurfer on that given day; thus it is worth applying some generally accepted model selection procedure in order to choose the model with the best combination of influences. We interpret any mechanism not selected through this procedure as one that is not influential in this process (i.e., we reject the hypothesis that the mechanism is a substantial factor in shaping the evolution of this network, net of other mechanisms). In this particular case we perform model selection using the BIC score, selecting the model in which the BIC is lowest (it may be seen that the full model is the best fitting model under this criterion, see Table~\ref{tbl:vertex} or Table~\ref{tbl:edge} and therefore each mechanism proposed is tested directly).

Overall, we find that the best-fitting model for the vertex process is one that incorporates differential base rates of attendance for ``regulars'' and (above and beyond this) for members of Group 1, as well as simple inertia, prior participation in cohesive conversation subgroups, and weekly seasonality. Thus, the data support the contention that all of the conjectured mechanisms for the attendance process are active in this case.  For the edge process, we likewise find that all conjectured mechanisms -- mixing, individual heterogeneity, contagious participation, inertia, prior embeddedness, and seasonality -- are active in governing who communicates with whom (conditional on who shows up).  Interpretation of model parameters is discussed below.

\subsubsection{Model Adequacy}

To evaluate the model adequacy of the best fitting model, we employ simulation-based one-step prediction under a inhomogeneous Bernoulli classifier as discussed in Section~\ref{subsec:gof}.  While the selected model may be the best fitting of those available, we are also interested in assessing the extent to which it can effectively capture the properties of the evolving beach network per se; significant failures in this regard may suggest the need for for further elaboration.  In the present case, we begin with simple network features such size and density (and, therefore, mean degree). In the context of interpersonal communication on a beach, capturing local group structure is also of interest; thus we include the statistics of the undirected triad census (null, dyad, two path, and triangle) as targets for evaluation\footnote{It is known that the triad census governs a number of key network statistics, such as transitivity; see also \citet{faust10}.}. To evaluate our ability to capture inequality in communication, we include degree centralization \citep{freeman:1979a}.  And, lastly, we may be interested in our ability to predict the extent to which the communication network formed on a given day will be well-connected, a feature that we examine using the Krackhardt connectedness statistic \citep{krackhardt94}. The simulation intervals for each GLI under the best fitting model (Model 4;  Figure~\ref{fig:gof2}) perform reasonably well under the criterion suggested in Section \ref{subsec:gof} ($\alpha \leq 0.95$). Under the 0.95 criterion, our model performs reasonably well; in Table~\ref{tbl:count} we see that the observed GLI falls within the interval over 26 of the 28 predicted time points for all but Mean Degree (and in fact falls within the interval all 28 times for 5 of 9 GLIs).

\begin{center}
[Table~\ref{tbl:count}]
\end{center}

\begin{center}
[Figure~\ref{fig:gof2}]
\end{center}

\begin{sidewaysfigure}
\begin{center}
	\textbf{5-Step Projection of \citeauthor{freeman88}'s 1986 Windsurfer Network}
   \includegraphics[width=1\linewidth]{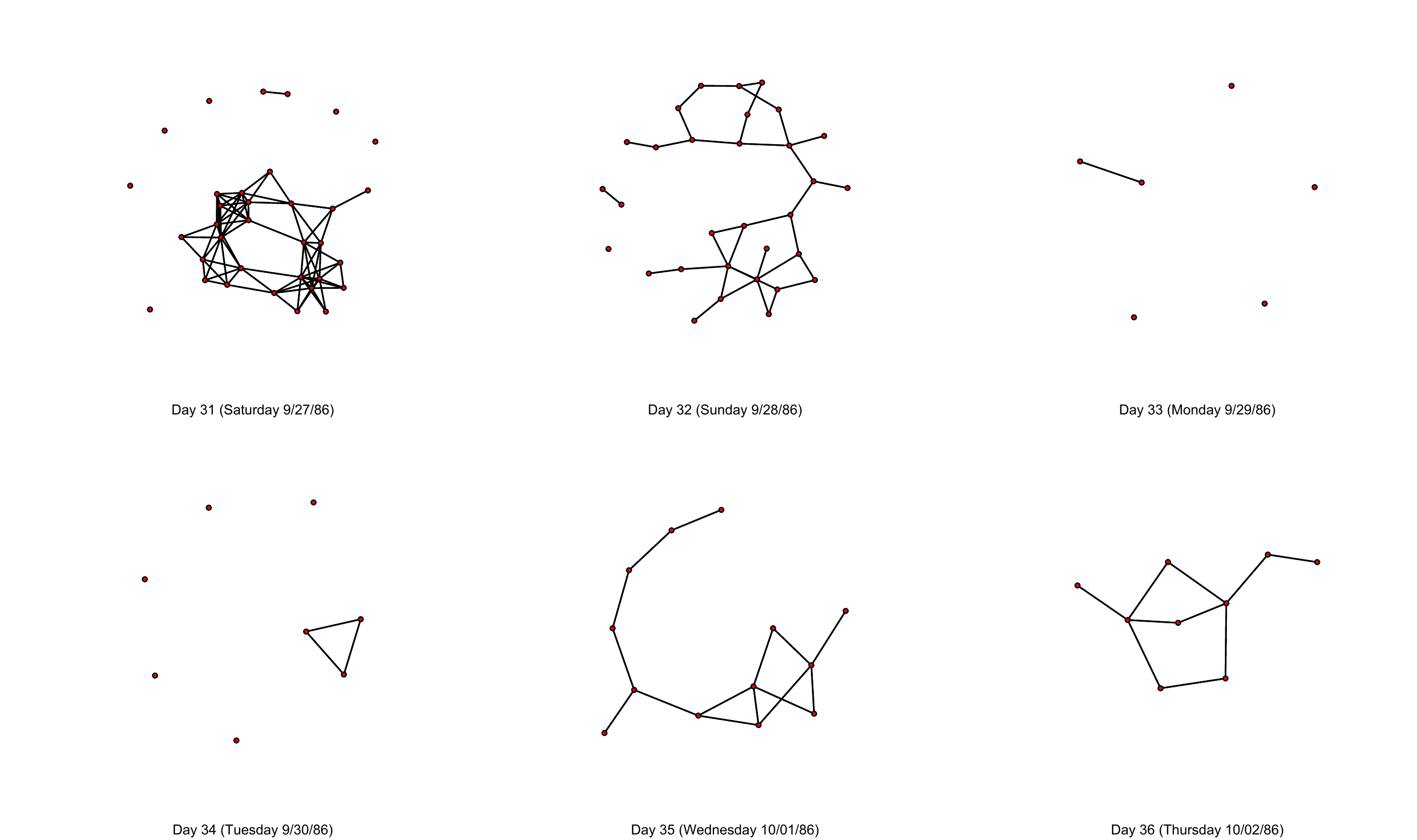} 
   \caption{The first of these six plots is the last observed network in \citeauthor{freeman88}'s \citeyear{freeman88} network (day 31). The next five graphs represent the 5 day projection under the inhomogenious Bernoulli classifier assumptions.}
   \label{fig:nstep}
   \end{center}
\end{sidewaysfigure}

Lastly we perform a 5-step prediction of the full model (Figure~\ref{fig:nstep}) as form of visual analysis to verify that the model is not producing degenerate structures; these could include, for example, those identified by \citet{robins05}, such as giant ``clumps,'' so called ``caveman'' graphs, or other highly clustered graphs. Such structures are largely considered pathological and unrepresentative of ``real-world" social networks, and (more importantly) do not resemble the types of networks arising within our observed data. Inspection of the graphs generated through the 5-step prediction verifies that the networks predicted by the model are non-pathological, either in terms of converging to an unrepresentative canonical structure (as in the Robins et al. case), or in producing effectively random graphs with less structure than the observed data.  Taken together with the GLI-based adequacy checks, these results suggest that the model is indeed doing a reasonable job of capturing the core features of the evolving network.

\begin{center}
[Figure~\ref{fig:nstep}]
\end{center}

\subsection{Parameter Interpretation}

The parameter estimates presented in Tables~\ref{tbl:vertex} and \ref{tbl:edge}  are interpreted in terms of the influence of the mechanisms proposed in Section~\ref{subsec:mech}. To simplify presentation, we discuss these mechanisms in two parts, starting with vertex mechanisms and proceeding to mechanisms associated with the edge set.

\subsection{Vertex Mechanisms}

We proposed three basic mechanisms for the vertex set dynamics in this particular context (Table~\ref{tbl:vertex}: Model 4). The first was whether or not an individual's group status was predictive of attendance.  As expected, we find that being a ``regular'' has a significant and positive influence over whether an individual is likely to appear on any given day (versus being an ``irregular''), with those in Group 1 being even more likely to appear. Mechanism two was that being present at the beach on the prior day before would make one more likely to appear at the beach on the next day, which is indeed what we find (the weight is again positive and significant). Similarly, if one is engaged in a conversational clique the day before then one is even more likely to appear the next day than if one is simply present; indeed, \emph{each} 3-clique in which an individual participates increases his or her conditional odds of subsequent attendance by over 40\%. Finally, we see that beach attendance is indeed highly seasonal: with the exception of a slight bump on Thursday, weekends are substantially more popular times for beach-going than the work week (Tuesdays, in particular).  These seasonal effects are comparable in magnitude to the effect of being a regular, and exceed the effect of inertia (although inertia combined with participation in 1--2 conversation clusters has a similar overall effect).

\begin{center}
[Table~\ref{tbl:vertex}]
\end{center}

\subsubsection{Edge Mechanisms}

We proposed five basic mechanisms shaping whether or not a beach goer was likely to engage in interpersonal communication (Table~\ref{tbl:edge}: Model 4), starting with assortative mixing of regulars (and group members). The mixing hypothesis is confirmed such that regulars are more likely to interact with other regulars, but refuted in the sense that irregulars are more likely to interact with regulars than with other irregulars. This suggests a core-periphery phenomenon, wherein irregulars are more likely to interact with ``core'' regulars who go to the beach more often and are more likely to be knowledgeable of the sport and area. Mechanism four, individual differences within the most influential group (Group 1) is confirmed: all but one individual is significantly more likely to interact or less likely to interact than baseline. This occurs at substantially high levels (as much as a plus 2.5 times or down to as low as negative 1.14 times). Mechanism three, contagious participation, is highly influential and is both positive and significant. The inertial hypothesis is confirmed since the lag and the cycle term are positive and significant (it should be pointed out that that the number of cycles cumulative up to 9 that a dyad may be involved in can be quite large (e.g. in the thousands) and thus this term can be quite influential). 

For mechanism five, it is important to point out that many of these terms cannot be interpreted independently. For example, everyone regardless of their categorization of ``regularity'' is influenced by the number of individuals on the beach on a given day. To put this in perspective, take the highest number of individuals to appear on the beach over the 31 days (37 individuals) so that $\log(37)\cdot 2.72 = 4.30$ and compare it to the lowest $\log(3)\cdot 2.72 = 2.99$. To fully grasp how this interacts with the days of week it is important to note that network size is highly correlated with the day of the week and thus we find that there are more individuals on the beach on a typical Saturday or Sunday than on a typical weekday (e.g., the lowest day occurs on a Wednesday and the highest day occurs on a Sunday), such that the total effect on baseline density at the high end is $\log(37)\cdot 2.72  - 1.69 = 8.13$ versus a total lowest day effect of $\log(3)\cdot 2.72 + 1.14 = 4.13$.  Thus the baseline propensity for interaction is given almost twice the boost (on logit scale) on the day with largest number of beach goers versus the day with the smallest number of beach goers. We therefore observe that, as the beach becomes more populated, the chance of interacting with any given individual increases, and therefore we find evidence for our hypothesis of contagious participation. 

\begin{center}
[Table~\ref{tbl:edge}]
\end{center}

Once the set of beach goers is chosen, the important factors which predict if any two or more individuals will interact stem from his or her ethnographically defined group (i.e., the ``regulars;'' this is especially true if he or she is part of Group 1) with the specific effect that all individuals regardless of status are likely to interact with the regular attendees. Individual differences in baseline propensity to interact are important, but only for Group 1 members, where this can be a quite large effect.  Thus, predicting which of the Group 1 members will appear on a given day is identified as an important factor in model success. The base activity in the network is greatly influenced by the number of individuals who appear on a given day, with a higher probability for interaction between every individual on the beach. Finally, an individual engaged in activity in the immediate past is more likely to engage in activity again in the present, and this effect is magnified if the individual is embedded in a larger conversational structures. 

\section{Discussion and Conclusion}
\label{set:conc}

The dynamic network logistic regression framework proposed in this article builds on a number of well-established concepts in the social network literature. We have extended this prior work by incorporating vital dynamics, clarifying the assumptions needed to model joint vertex/edge dynamics in logistic form, and addressing practical issues such as model assessment and scalability.  Applying the resulting framework to a dynamic network, we illustrated how this approach allows us to identify mechanisms underlying both individual presence/absence and relationships in a straightforward fashion.

Based on our model adequacy checks, we find that our proposed model does a reasonable job of capturing many properties of the beach data, despite the lack of available covariates (e.g., age, race, prior relationships) that would undoubtedly facilitate prediction.  Notwithstanding our model's limitations, we find that the mechanisms most important to prediction of dynamic network collaboration in the Southern California beach data are assortative mixing, inertia (in dyadic sense and in the number of cycles one is engaged in), individual differences of key players, the size of the network itself, and seasonality. As expected, we find that those identified ethnographically as core members of the beach community are more likely to be present on a given day, along with factors such as having been active on a previous day, and having been previously involved in group interaction.  We also find that the day of the week greatly influences the number of individuals who appear on any given day.

\begin{sidewaysfigure}
\begin{center}
   \includegraphics[width=1\linewidth]{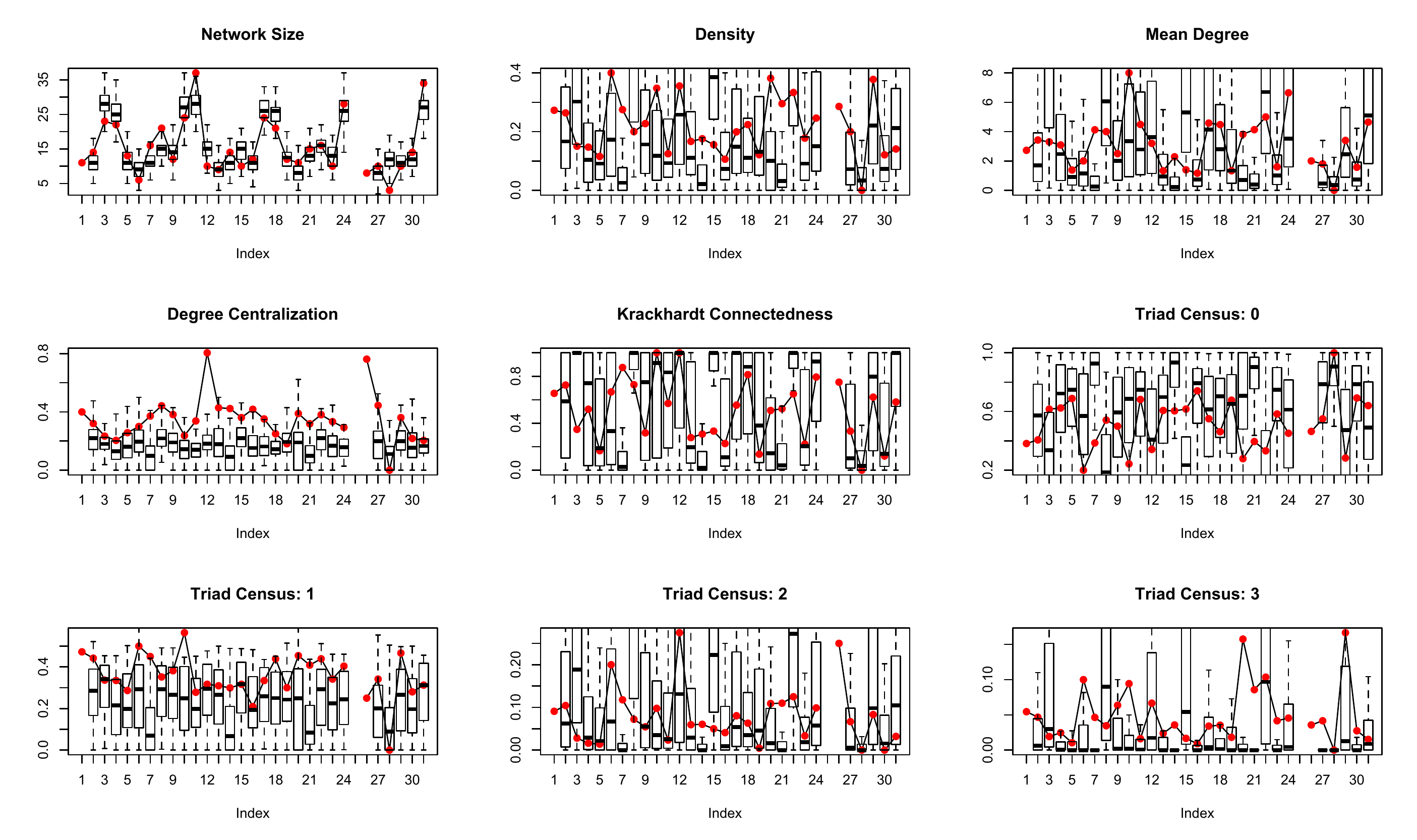} 
   \caption{ GLI  comparison for the one-step dynamic network logistic regression prediction under an inhomogeneous Bernoulli classifier assumption for Model 4. Red dots represent the observed GLI and black boxplots are the simulated distribution of the one-step prediction based on the model 4 weights (100 simulated networks for each one-step prediction).}
   \label{fig:gof2}
   \end{center}
\end{sidewaysfigure}

\begin{sidewaysfigure}
\begin{center}
   \includegraphics[width=1\linewidth]{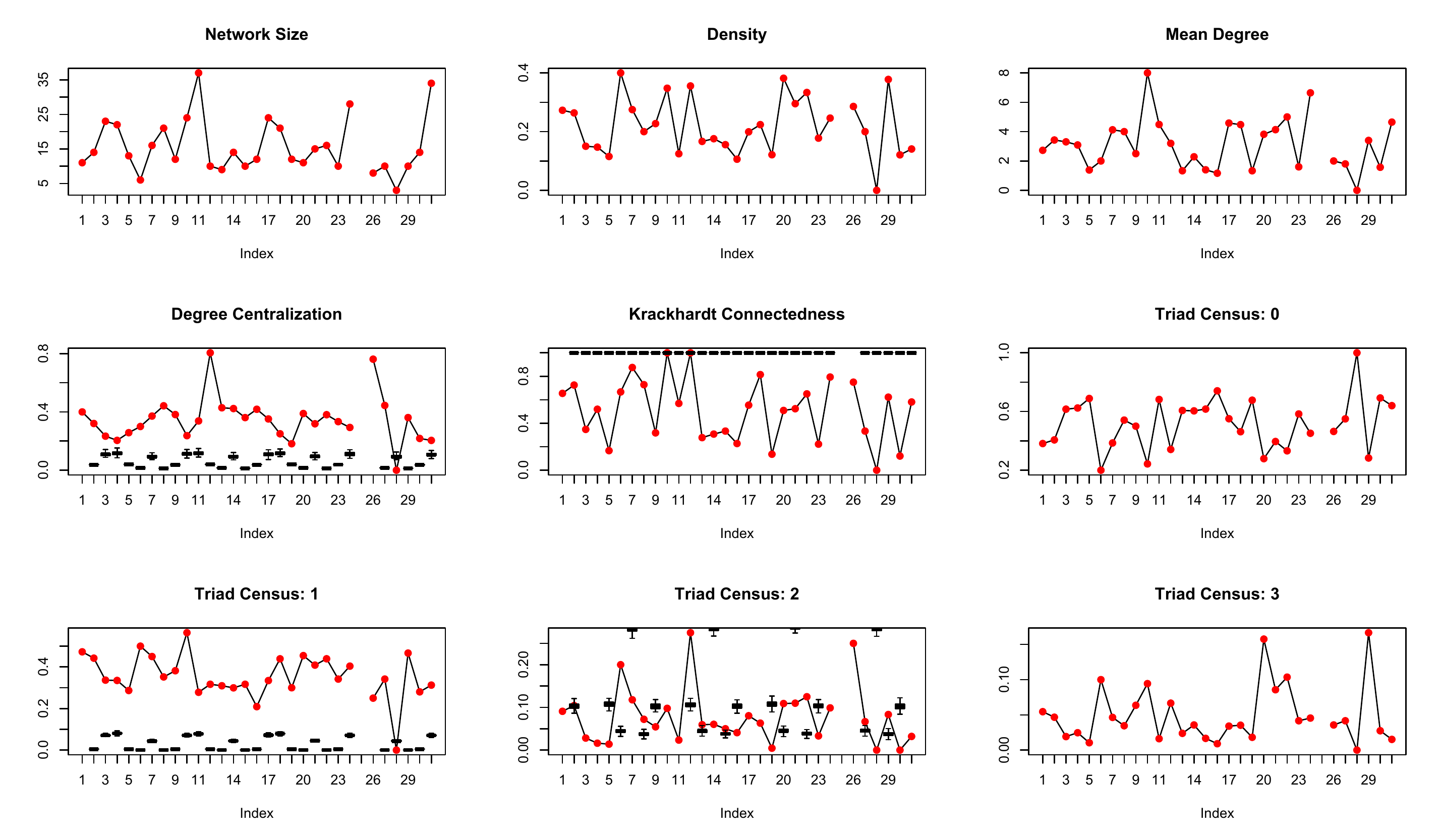} 
   \caption{ GLI  comparison for the one-step dynamic network logistic regression prediction under an inhomogeneous Bernoulli classifier with the vertex set fixed to $V_{max}=95$. Red dots represent the observed GLI and black boxplots are the simulated distribution of the one-step prediction based on the model 4 edge weights (100 simulated networks for each one-step prediction).}
   \label{fig:gof1}
   \end{center}
\end{sidewaysfigure}

We have noted repeatedly throughout the paper that a good vertex set model is key to effective prediction of joint vertex/edge set evolution, a fact that can be dramatically illustrated by comparing the model of Section~\ref{subsec:gof2} to a similar model for which the vertex set is fixed to $V_{max}$ (i.e., assuming all actors are eligible to interact) and the best edge model.  The results are shown in Figure~\ref{fig:gof1}. Notice that the model simulation intervals never cover the observed statistics, and are often so far from the observed values that they do not fall within the range of the observed statistics over the entire observation period (Figure~\ref{fig:gof2}). A naive approaches to solving the vertex problem clearly will not work in this context.

\begin{center}
[Figure~\ref{fig:gof1}]
\end{center}

Comparing performance of our best-fit model to a naive model without a well-specified vertex component underscores the critical interaction between the size and composition of the vertex set and the structure of the resulting relationships.  In particular, we find that models that do not accurately capture vertex set dynamics are deeply pathological for predicting other aspects of structure as well: one simply cannot get the edge set right without first modeling the vertex set.  Since vertex set models are rarely employed at present, this observation calls into question the trustworthiness of the current generation of dynamic network models.  While more research is certainly needed on this point, our experience thus far has strongly suggested that predictive adequacy for dynamic network models in realistic settings will depend as much or more heavily on capturing the factors that lead to individual presence and participation than on modeling the factors that lead participating individuals to interact.  This implies a substantial rethink of our current ideas regarding network evolution.

Although we believe that the logistic framework pursued here is both flexible and powerful, we wish to end on a note of moderation.  There may well be settings for which the available historical data does not adequately account for dependence among edges (or vertices), and for which the logistic approximation will perform poorly.  Likewise, some research questions may require a degree of predictive accuracy that cannot be readily obtained without incorporating simultaneous dependence.  For these problems, the framework presented here should be viewed as a ``first cut'' family of models, to be extended by the incorporation of additional dependence terms in a manner analogous to the extension of Bernoulli graph models in the cross-sectional ERGM case.  That said, considerable progress may be made by beginning one's investigation with models based on conditional independence assumptions, and adding dependence terms only as needed to obtain acceptable results.  Since the dynamic logistic models can be easily manipulated (and understood), they are well-suited to exploratory analysis, and to tasks such as the identification of key covariates.  They also scale readily to large data sets, making them applicable in settings for which models with edgewise dependence are too computationally expensive to be employed.  These advantages make the dynamic logistic family an important and useful tool in the analyst's arsenal, as part of the growing family of techniques for modeling the dynamics of social structure.


\bibliographystyle{elsarticle-harv}
\bibliography{dynlogit}
\clearpage


\begin{algorithm}[h!] 
\caption{Inhomogeneous Bernoulli classifier}\label{alg:bc} 
\begin{algorithmic}[1] 
\FOR{$i=1$ to m}
\FOR{$t=k$ to $T-1$}
\STATE $\hat{\pi}_{t+1}^v =$ Predicted vertex probabilities from model
\STATE p = 0
\FOR{ $l=1$ to $n$}
\IF{Bernoulli($\hat{\pi}_{t+1,l}^v) == 1$}
\STATE $\hat{V}_{t+1}[p] = v_{l}$
\STATE p = p+1
\ENDIF
\ENDFOR
\STATE $\hat{\pi}_{t+1}^e|\hat{V}_{t+1}$ = Predicted edge probabilities from model
\STATE $\hat{Y}_{t+1} =$ Bernoulligraph$(\hat{\pi}_{t+1}^e|\hat{V}_{t+1})$
\STATE Save[i,t] = $GLI(\hat{Y}_{t+1})$
\ENDFOR
\ENDFOR
\STATE
\COMMENT{$T$ = number of time points, $GLI(\cdot)$ is function which returns a GLI or a vector of GLIs, Save is an $m$ by $t$ matrix, and $\pi_{t+1}^{\cdot}$ represent the predicted probabilities, where $v$ denotes the predicted probabilities of the vertex set and $e$ denotes the predicted probabilities of the edge set.}

\end{algorithmic} 
\end{algorithm} 

\clearpage


\begin{table}[ht]
\begin{center}
\begin{tabular}{l|c}
 \multicolumn{2}{c}{\textbf{GLI One-Step Prediction Simulation Count ($\alpha \leq 0.95$)}} \\
  \hline
GLI & \# Correct\\
  \hline
Network Size & 26 \\ 
  Density & 28 \\ 
  Mean Degree & 28 \\ 
  Degree Centralization & 20\\ 
  Krackhardt Connectedness & 28 \\ 
  Triad Census: 0 & 28 \\ 
  Triad Census: 1 & 26 \\ 
  Triad Census: 2 & 28 \\ 
  Triad Census: 3 & 27\\ 
   \hline
\end{tabular}
\caption{Check of whether the $\alpha \leq 0.95$ simulation interval contains a given GLI. Total possible correct is 28.}\label{tbl:count}
\end{center}
\end{table}

\begin{sidewaystable}
\begin{center}
\footnotesize{
\begin{tabular}{lcccccccc}
   \multicolumn{9}{c}{\textbf{Vertex Model}} \\
   \hline
   \hline
 & Model 1 & s.e.  & Model 2 & s.e.  & Model 3 & s.e. & Model 4 & s.e.  \\ 
  \hline
  \hline
\multicolumn{1}{c}{BIC}  & 8166.2395 &  & 7929.2997 &  & 7727.7134 &  & 7695.6264 &  \\ 
\hline
\hline
  Intercept & -1.5893* & (0.0517) & -1.9250* & (0.0536) & -2.5564* & (0.0550) & -3.2988* & (0.0567) \\ 
  \hline
  $\mathbb{I}$\{Regular\} &  &  &  &  &  &  & \ 0.9245* & (0.0649) \\ 
  $\mathbb{I}$\{Group 1\} &  &  &  &  &  &  & \ 0.7827* & (0.0912) \\ 
  \hline
  $V_{t-1}$ &  &  & 1.5295* & (0.1010) & 1.1195* & (0.1061) &\ 0.7764* & (0.1100) \\ 
  3-Cycle$_{t-1}$ &  &  &  &  & 0.3701* & (0.0598) & \ 0.3452* & (0.0615) \\ 
  \hline
    $\mathbb{I}$\{Monday\} &  &  &  &  & -- &  & -- &  \\ 
  $\mathbb{I}$\{Tuesday\} &  &  &  &  & 0.0421 \ & (0.1799) & -0.0827\ & (0.1839) \\ 
  $\mathbb{I}$\{Wednesday\} &  &  &  &  & 0.3919* & (0.1628) & \ 0.2713 \ & (0.1647) \\ 
   $\mathbb{I}$\{Thursday\}  &  &  &  &  & 0.5009* & (0.1500) & \ 0.4182* & (0.1556) \\ 
  $\mathbb{I}$\{Friday\}&  &  &  &  & 0.2788 \ & (0.1445) & \ 0.2014 \ & (0.1469) \\ 
   $\mathbb{I}$\{Saturday\}  &  &  &  &  & 1.3813* & (0.1059) & \ 1.3719* & (0.1115) \\ 
  $\mathbb{I}$\{Sunday\} &  &  &  &  & 1.0332* & (0.1445) & \ 1.0739* & (0.1479) \\ 
   \hline
\end{tabular}
\caption{Vertex portion of Models 1 through 4 ranked by BIC score. Significance: `${}^*$' p-value $<$ 0.05. $t-1$ indicates the lag interval; $V_{t-1}$ represents the lag term;  3-Cycle$_{t-1}$ represents the triadic effect; and $\mathbb{I}$ indicates that a given variable is represented via dummy variable.}\label{tbl:vertex} 
}
\end{center}
\end{sidewaystable}

\begin{sidewaystable}
\begin{center}
\footnotesize{
\begin{tabular}{lcccccccc}
\multicolumn{9}{c}{\textbf{Edge Model}} \\
  \hline
 & Model 1 & s.e. & Model 2 & s.e. & Model 3 & s.e. & Model 4 & s.e. \\ 
  \hline
  \hline
\multicolumn{1}{c}{BIC}  & 8166.2395 &  & 7929.2997 &  & 7727.7134 &  & 7695.6264 &  \\ 
\hline
\hline
 Intercept & -2.3162* & (0.0359) & -4.3081* & (0.0362) &  &  &  &  \\ 
 \hline
  Mixing\{Regular (R)\} &  &  &  &  & -4.8336* & (0.0430) & -8.2250* & (0.0434) \\ 
  Mixing\{$\neg$ R\}   &  &  &  &  & -6.0214* & (0.2481) & -9.4768* & (0.2517) \\ 
  Mixing\{$\neg$ R  $\leftrightarrow$ R\}  &  &  &  &  & -5.3853* & (0.0736) & -8.8148* & (0.0741) \\ 
\hline
  $\mathbb{I}$\{Indiv \ 1 (G1)\}  &  &  &  &  &  &  & -0.2999* & (0.0551) \\ 
  $\mathbb{I}$\{Indiv \ 2 (G1)\}  &  &  &  &  &  &  & -1.1425* & (0.0624) \\ 
  $\mathbb{I}$\{Indiv \ 3 (G1)\}  &  &  &  &  &  &  & \ 2.5173* & (0.0768) \\ 
  $\mathbb{I}$\{Indiv \ 4 (G1)\}  &  &  &  &  &  &  & -0.8367* & (0.0730) \\ 
  $\mathbb{I}$\{Indiv \ 5 (G1)\}  &  &  &  &  &  &  & -0.4407* & (0.0485) \\ 
  $\mathbb{I}$\{Indiv \ 6 (G1)\}  &  &  &  &  &  &  & -1.0009* & (0.0384) \\ 
  $\mathbb{I}$\{Indiv \ 7 (G1)\}  &  &  &  &  &  &  & -0.3989* & (0.0476) \\ 
  $\mathbb{I}$\{Indiv \ 8 (G1)\}  &  &  &  &  &  &  & \ 0.8282* & (0.0570) \\ 
  $\mathbb{I}$\{Indiv \ 13 (G1)\}  &  &  &  &  &  &  & -0.3263* & (0.0916) \\ 
  $\mathbb{I}$\{Indiv \ 16 (G1)\}  &  &  &  &  &  &  & -0.2785* & (0.0473) \\ 
  $\mathbb{I}$\{Indiv \ 17 (G1)\}  &  &  &  &  &  &  & 0.4985* & (0.0397) \\ 
  $\mathbb{I}$\{Indiv \ 18 (G1)\}  &  &  &  &  &  &  & -0.6768* & (0.0756) \\ 
  $\mathbb{I}$\{Indiv \ 19 (G1)\}  &  &  &  &  &  &  & \ 0.5061* & (0.0580) \\ 
  $\mathbb{I}$\{Indiv \ 22 (G1)\}  &  &  &  &  &  &  & -0.8799* & (0.0816) \\ 
  $\mathbb{I}$\{Indiv \ 28 (G1)\}  &  &  &  &  &  &  & -0.3846* & (0.0504) \\ 
  $\mathbb{I}$\{Indiv \ 37 (G1)\}  &  &  &  &  &  &  & \ 0.1560* & (0.0459) \\ 
  $\mathbb{I}$\{Indiv \ 39 (G1)\}  &  &  &  &  &  &  & -0.7093* & (0.0637) \\ 
  $\mathbb{I}$\{Indiv \ 40 (G1)\}  &  &  &  &  &  &  & \ 0.0575 \ & (0.0552) \\ 
  $\mathbb{I}$\{Indiv \ 46 (G1)\} &  &  &  &  &  &  & -0.3053* & (0.0501) \\ 
  $\mathbb{I}$\{Indiv \ 47 (G1)\}  &  &  &  &  &  &  & 1.4813* & (0.1262) \\ 
  $\mathbb{I}$\{Indiv \ 51 (G1)\}  &  &  &  &  &  &  & 0.9822* & (0.0713) \\ 
  $\mathbb{I}$\{Indiv \ 54 (G1)\}  &  &  &  &  &  &  & 0.2812* & (0.0645) \\ 
  \hline
     $\log(n_t )$  &  &  & 0.6394* & (0.0118) & 0.4291* & (0.0120) & 2.7223* & (0.0121) \\ 
    \hline
  $Y_{t-1}$ &  &  & 0.8946* & (0.1023) & 0.3169* & (0.1040) & 0.2861* & (0.1053) \\ 
  $\log($9-Cycle$_{t-1}+1)$ &  &  &  &  & 0.0860* & (0.0094) & 0.1041* & (0.0095) \\ 
  \hline
   $\mathbb{I}$\{Monday\} &  &  &  &  & -- &  & -- &  \\ 
  $\mathbb{I}$\{Tuesday\} &  &  &  &  & 1.6966* & (0.1649) & \ 0.9565* & (0.1668) \\ 
   $\mathbb{I}$\{Wednesday\} &  &  &  &  & 2.0953* & (0.1246) & -1.2812* & (0.1255) \\ 
  $\mathbb{I}$\{Thursday\}  &  &  &  &  & 1.7549* & (0.1060) & \ 1.1441* & (0.1079) \\ 
 $\mathbb{I}$\{Friday\} &  &  &  &  & 1.1324* & (0.1304) & \ 0.0628 \ & (0.1311) \\ 
  $\mathbb{I}$\{Saturday\}  &  &  &  &  & 1.5712* & (0.0566) & -1.6043* & (0.0569) \\ 
   $\mathbb{I}$\{Sunday\}&  &  &  &  & 0.8772* & (0.0838) & -1.6944* & (0.0840) \\ 
  \hline
  \end{tabular}
\caption{Edge portion of Models 1 - 4 ranked by BIC score. Significance: `${}^*$' p-value $<$ 0.05. lag is $t-1$; Note that the mixing term stands in for the intercept term in Models 3 and 4; Indiv $k$ indicates the density effect for individual $k$ (as indexed in the data set); $\log(n_t$ indicates the contagious propensity effect and is the log of the network size at time $t$; again, $Y_{t-1}$ represents the lag effect and $\log($9-Cycle$_{t-1}+1)$ is the embeddeness cycle statistic; and $\mathbb{I}$ indicates that a given variable is represented via dummy variable.}\label{tbl:edge}
}
\end{center}
\end{sidewaystable}


\end{document}